\documentclass[conference]{IEEEtran}
\IEEEoverridecommandlockouts
\usepackage{bm}
\usepackage{cite}
\usepackage{amsmath,amssymb,amsfonts}
\usepackage{algorithm}
\usepackage{algpseudocode}
\usepackage{graphicx}
\usepackage{textcomp}
\usepackage{xcolor}
\usepackage{subfigure}

\def\BibTeX{{\rm B\kern-.05em{\sc i\kern-.025em b}\kern-.08emT\kern-.1667em\lower.7ex\hbox{E}\kern-.125emX}}
\newtheorem{assumption}{Assumption}

\newtheorem{theorem}{Theorem}

\begin{document}
\title{Federated Learning via Active RIS Assisted Over-the-Air Computation}
\author{\IEEEauthorblockN{Deyou Zhang*, Ming Xiao*, Mikael Skoglund*, and H. Vincent Poor$^\S$}
\IEEEauthorblockA{\textit{*Division of Information Science and Engineering, KTH Royal Institute of Technology, Stockholm 10044, Sweden} \\
\textit{$^\S$Department of Electrical and Computer
Engineering, Princeton University, Princeton, NJ 08544, USA} \\
email: \{deyou, mingx, skoglund\}@kth.se; poor@princeton.edu}
}

\maketitle

\begin{abstract}
In this paper, we propose leveraging the active reconfigurable intelligence surface (RIS) to support reliable gradient aggregation for over-the-air computation (AirComp) enabled federated learning (FL) systems. An analysis of the FL convergence property reveals that minimizing gradient aggregation errors in each training round is crucial for narrowing the convergence gap. As such, we formulate an optimization problem, aiming to minimize these errors by jointly optimizing the transceiver design and RIS configuration. To handle the formulated highly non-convex problem, we devise a two-layer alternative optimization framework to decompose it into several convex subproblems, each solvable optimally. Simulation results demonstrate the superiority of the active RIS in reducing gradient aggregation errors compared to its passive counterpart.
\end{abstract}

\begin{IEEEkeywords}
Federated learning, over-the-air, reconfigurable intelligent surface, active RIS.
\end{IEEEkeywords}

\IEEEpeerreviewmaketitle

\section{Introduction}
Federated learning (FL) has emerged as a promising distributed machine learning alternative to centralized learning approaches \cite{FL-Google}. Orchestrated by an edge server, FL enables multiple edge nodes to collaboratively train a shared model without directly unveiling raw data. Specifically, FL operates in an iterative manner consisting of two primary steps: 1) The edge server disseminates a global model parameter vector to edge nodes for distributed on-device training with their local data. 2) These edge nodes upload their locally computed model parameter vectors to the edge server to update the global model parameter vector as a weighted average of the local vectors. Since only model parameter vectors rather than raw data are aggregated at the edge server, FL avoids prohibitive data transmission delay and mitigates potential privacy disclosure.

Despite the considerable advantages of FL, uploading local model parameter vectors to the edge server via conventional orthogonal multiple access (OMA) schemes can be resource-intensive, emerging as a potential bottleneck in FL. Though a number of works have proposed optimizing the communication and computation resources of edge nodes to enhance model uploading efficiency \cite{MingzheChen-TWC, ZhaohuiYang-TWC, HaoChen-IoTJ}, they did not exploit the waveform-superposition property of the multiple-access channel, thus not fully harnessing the benefits of wireless communications. As an alternative, over-the-air computation (AirComp) has been recently introduced to enable simultaneous local model uploading over shared radio resources \cite{ZhibinWang-OTA}.

Unlike OMA schemes that allocate orthogonal radio resources such as time or bandwidth to edge nodes for independent transmission, the radio resources required by AirComp-enabled model uploading are independent of the number of edge nodes, significantly enhancing communication efficiency and system scalability. Despite these advantages, AirComp-enabled FL suffers from model aggregation errors caused by wireless fading and noise \cite{GuangxuZhu-TWC, YuanmingShi-TWC, YongmingHuang-JSAC, ShuaiWang-TWC, Deniz-TSP}. Existing works in this area avoided large aggregation errors mainly by excluding ``stragglers'', i.e., devices with weak channels, from concurrent model uploading \cite{GuangxuZhu-TWC, YuanmingShi-TWC}. For example, the authors in \cite{GuangxuZhu-TWC} proposed a truncated-based power control scheme to discard devices in deep fading. However, discarding devices from training reduces the number of training data, inevitably compromising learning performance, especially when the discarded devices possess unique data samples.

To cope with the straggler issue and mitigate potential degradation in learning performance caused by large aggregation errors, an alternative strategy is to strengthen the communication channels between stragglers and the edge server using advanced communication technologies, such as relays \cite{HangLiu-Relay} or passive reconfigurable intelligent surfaces (RISs) \cite{HangLiu-TWC, ZhibinWang-TWC, ShaochengHuang-RIS}. For instance, the authors in \cite{HangLiu-TWC} proposed to jointly optimize device selection, transceiver design, and RIS configuration to partially alleviate the straggler issue in FL. Though with some merits, the ``multiplicative fading'' effect curtails the benefits of passive RISs. To overcome such a fundamental limitation of passive RISs, a novel RIS architecture named active RIS has appeared \cite{YCLiang-ActiveRIS, LinglongDai-ActiveRIS}. Unlike its passive counterpart, the active RIS is capable of amplifying its reflected signals through integrated reflection-type amplifiers in its reflecting elements.

In this paper, we focus on the AirComp-enabled FL system and propose utilizing active RIS to control gradient aggregation errors during training. Firstly, we analyze the convergence property of the considered FL system and derive an upper bound on the expected difference between the training loss and the optimal loss. This analysis reveals that minimizing the mean squared error (MSE) between the target global gradient vector and the received one is crucial for narrowing the convergence gap. Subsequently, we aim to minimize the MSE by jointly optimizing the transceiver design and RIS configuration. To address such a highly non-convex optimization problem, we introduce a two-layer alternative optimization (AO) strategy to decompose the original problem into several convex subproblems, each optimally solvable. Finally, we employ the MNIST dataset to assess learning performance in the context of the handwritten digit recognition task. Experiment results demonstrate the superiority of the active RIS in reducing gradient aggregation errors compared to its passive counterpart.

Throughout this paper, we use regular, bold lowercase, and bold uppercase letters to denote scalars, vectors, and matrices, respectively; $\mathcal R$ and $\mathcal C$ to denote the real and complex number sets, respectively; $(\cdot)^T$ and $(\cdot)^H$ to denote the transpose and the conjugate transpose, respectively. We use $x$ to denote a typical entry of $\bm x$; $\|\bm x\|$ to denote the $\ell_2$-norm of $\bm x$; ${\rm diag}(\bm x)$ to denote a diagonal matrix with its diagonal entries specified by $\bm x$; $|\mathcal D|$ to denote the cardinality of set $\mathcal D$. We use $\bm I$ to denote the identity matrix; $\mathcal {CN}(\bm \mu, \bm \Sigma)$ to denote the complex Gaussian distribution with mean $\bm \mu$ and covariance matrix $\bm \Sigma$; $\nabla$ to denote the gradient operator, and ${\mathbb E}$ to denote the expectation operator.

\begin{figure}
\centering
\includegraphics[width = 8.6cm]{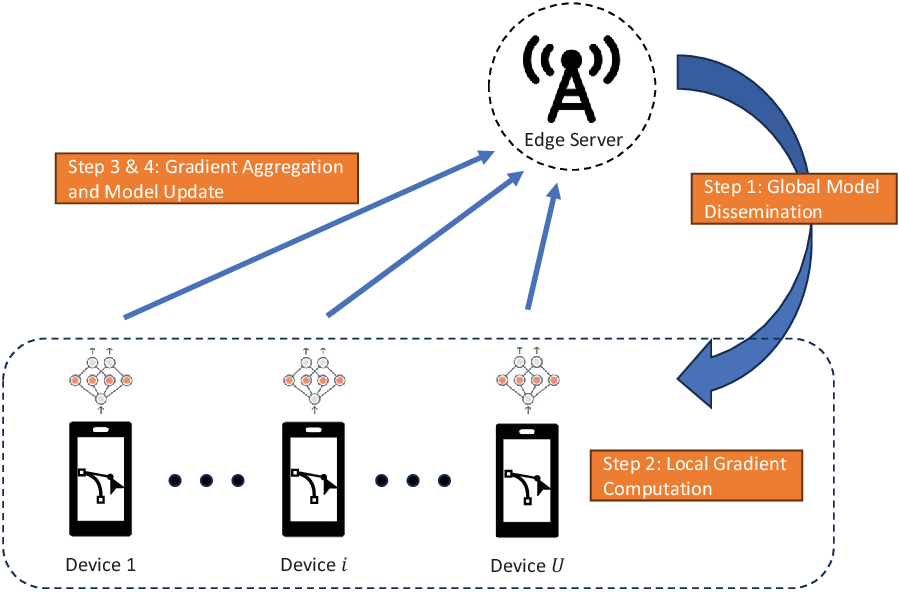}
\caption{A typical wireless FL system consisting of one edge server and multiple edge nodes.}\label{FL-Model}
\end{figure}

\section{System Model}\label{Section-SystemModel}
\subsection{FL Model}\label{Section-FLPreliminary}
The canonical FL system consists of an edge server and $U$ edge nodes, as shown in Fig. \ref{FL-Model}. By denoting the dataset and model parameter vector at edge node $i$, $\forall i \in \mathcal U \triangleq \{1, \cdots, U\}$, as $\mathcal K_i$ and $\bm w_i \in \mathcal R^{d \times 1}$, respectively, we can express the aim of FL using the following optimization problem:
\begin{subequations}\label{Eq-FL-Modeling}
    \begin{align}
    \min\limits_{\bm w_1, \cdots, \bm w_U, \bm w}~ & \frac{1}{\sum\nolimits_{j=1}^U K_j} \sum\limits_{i=1}^U \sum\limits_{k=1}^{K_i} \ell(\bm w_i, \bm u_{ik}, v_{ik}) \label{Eq-FL-Modeling-Obj} \\[2ex]
    \text{s.t.}~~~~~~ & \bm w_1 = \cdots = \bm w_U = \bm w, \label{Eq-FL-Modeling-Cons}
\end{align}
\end{subequations}
where $K_i = |\mathcal K_i|$ denotes the size of data samples in edge node $i$, $(\bm u_{ik}, v_{ik})$ denotes the $k$-th data sample in $\mathcal K_i$, $\ell(\bm w_i, \bm u_{ik}, v_{ik})$ is the loss function with respect to $(\bm u_{ik}, v_{ik})$, and $\bm w$ is often termed the global model parameter vector. Note that the objective function in \eqref{Eq-FL-Modeling-Obj} can be rewritten into a separable form
\begin{align}
    L(\bm w) & \triangleq \frac{1}{\sum\nolimits_{j=1}^U K_j} \sum\limits_{i=1}^U \sum\limits_{k=1}^{K_i} \ell(\bm w_i, \bm u_{ik}, v_{ik}) \nonumber \\[2ex]
    & = \frac{1}{\sum\nolimits_{j=1}^U K_j} \sum\limits_{i=1}^U K_i L_i(\bm w_i),
\end{align}
where $L_i(\bm w_i)$ is given by
\begin{equation}
    L_i(\bm w_i) = \frac{1}{K_i}\sum\limits_{k=1}^{K_i} \ell(\bm w_i, \bm u_{ik}, v_{ik}).
\end{equation}
As a result, the training of FL model parameters, i.e., solving \eqref{Eq-FL-Modeling}, can be implemented in a distributed and iterative manner, where the $t$-th iteration, also known as the training round, consists of the following steps.

\textbf{Global model dissemination}: The edge server disseminates the current global model parameter vector $\bm w^{[t]}$ to the $K$ edge nodes.

\textbf{Local gradient computation}: Upon receiving $\bm w^{[t]}$, each edge node uses its own dataset to compute a local gradient vector:
\begin{align}\label{Eq-LG}
    {\bm g}_i^{[t]} & \triangleq \nabla L_i(\bm w^{[t]}) \nonumber \\[2ex]
    & = \frac{1}{K_i}\sum\limits_{k=1}^{K_i} \nabla \ell(\bm w_i, \bm u_{ik}, v_{ik}),~\forall i \in \mathcal U.
\end{align}

\textbf{Gradient aggregation}: The $K$ edge nodes upload their respectively computed local gradient vectors to the edge server, which takes a weighted average of these local gradient vectors to get the global gradient vector \cite{HangLiu-TWC}:
\begin{equation}\label{Eq-MA}
    \bm g^{[t]} = \frac{1}{\sum\nolimits_{j=1}^U K_j} \sum\limits_{i = 1}^U K_i {\bm g}_i^{[t]}.
\end{equation}

\textbf{Global model update}: Once obtaining $\bm g^{[t]}$, we update the global model parameter vector by
\begin{equation}\label{Eq-MB}
    \bm w^{[t+1]} = \bm w^{[t]} - \eta^{[t]} {\bm g}^{[t]},
\end{equation}
where $\eta^{[t]} \ll 1$ is the learning rate.

Such a procedure is repeated for a maximum number of $T$ rounds or until the global consensus, i.e., \eqref{Eq-FL-Modeling-Cons}, is achieved.

\subsection{Active RIS with SI}
\begin{figure}
\vskip3pt
\centering
\includegraphics[width = 7.6cm]{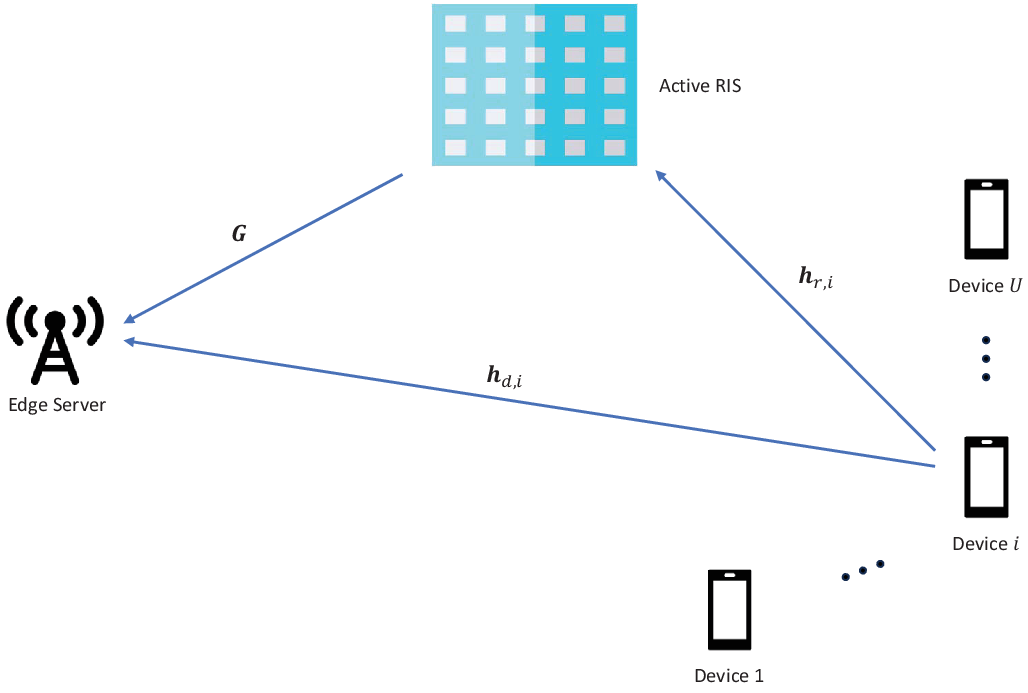}
\caption{The RIS assisted communication system.}\label{RIS-Model}
\end{figure}

To improve the channel quality between the edge server and the $U$ edge nodes, we propose deploying an $N$-element active RIS in the wireless FL system, as shown in Fig. \ref{RIS-Model}. As such, the equivalent channel between the edge server and each edge node now consists of three links, i.e., the device-server link, the device-RIS link, and the RIS-server link.

Moreover, since the RIS works in full-duplex mode, the self-interference (SI) occurs. By denoting the signal impinging on the active RIS as $\bm x_{\rm in}$, the reflected signal of the active RIS in the presence of SI, $\bm x_{\rm out}$, can be modeled as follows \cite{LinglongDai-ActiveRIS}
\begin{equation}\label{Eq-RIS-Model}
    \bm x_{\rm out} = (\bm I - \bm \Phi \bm H)^{-1} \bm \Phi \left(\bm x_{\rm in} + \bm z_A\right).
\end{equation}
In \eqref{Eq-RIS-Model}, $\bm \Phi = {\rm diag}\left(\beta_1 e^{j \theta_1}, \cdots, \beta_N e^{j \theta_N}\right)$ is the reflection coefficient matrix of the active RIS, where $\beta_n$ and $\theta_n$ respectively denote the amplification factor and phase shift of the $n$-th RIS element. It is worth mentioning that $\beta_n$ can be larger than one due to the integrated reflection-type amplifier in active RISs. In addition, $\bm H \in \mathcal C^{N \times N}$ in \eqref{Eq-RIS-Model} is the SI channel, and $\bm z_A \sim \left(\bm 0, \sigma_A^2 \bm I \right)$ is the thermal noise introduced at the active RIS. Regarding $\bm H$, we assume each of its elements follows ${\cal CN}(0, \nu^2)$, and when all of its elements are small, we can approximate \eqref{Eq-RIS-Model} as follows
\begin{equation}\label{Eq-RIS-Model2}
    \bm x_{\rm out} \approx (\bm I + \bm \Phi \bm H) \bm \Phi \left(\bm x_{\rm in} + \bm z_A\right).
\end{equation}

\subsection{AirComp-Enabled Gradient Aggregation}
As mentioned earlier, to reduce the communication overhead, we adopt AirComp for gradient aggregation. That is, the $U$ edge nodes transmit their respective local gradient vectors to the edge server using the same time-frequency resources at each training round. In the following, we will elaborate on the details.

First of all, each edge node normalizes its computed local gradient vector via
\begin{equation}\label{Eq-Gradient-Normalization}
    \bm s_i = \frac{\bm g_i - \bar g_i}{\delta_i},~\forall i \in \mathcal U,
\end{equation}
where $\bar g_i$ and $\delta^2_i$ denote the first-order and second-order statistics of $\bm g_i$, respectively. Note that in \eqref{Eq-Gradient-Normalization}, we have omitted the training round index $t$ for brevity. Via \eqref{Eq-Gradient-Normalization}, $\bm g_i$, $\forall i \in \mathcal U$, is normalized as a zero-mean and unit-variance vector $\bm s_i$, which is the information sequence sent by node $i$ for gradient aggregation\footnote{Alternatively, we can convert $\bm s_i$ into a $d/2$-length complex vector via $[\bm s_i]_{1:d/2} + {\rm j} [\bm s_i]_{d/2+1:d}$ for more efficient transmission.}. Recall that our target variable is $\bm g$, which can be rewritten as follows:
\begin{align}\label{Eq-MA2}
    \bm g = \frac{1}{K} \sum\limits_{i = 1}^U K_i {\bm g}_i = \frac{1}{K} \sum\limits_{i=1}^U K_i(\delta_i \bm s_i + \bar g_i),
\end{align}
where $K = \sum\nolimits_{i=1}^U K_i$. According to \eqref{Eq-MA2}, in order to obtain $\bm g$, we first need to obtain
\begin{equation}\label{Eq-MA3}
    \bm s \triangleq \sum\limits_{i=1}^U K_i \delta_i \bm s_i,
\end{equation}
which is a nomographic function of $\{\bm s_i\}$ and can be obtained via AirComp \cite{ZhibinWang-OTA}, as detailed below.

Let $s_i$ denote a typical entry of $\bm s_i$, and $\bm h_{r,i} \in \mathcal C^{N \times 1}$ denote the channel from edge node $i$ to the active RIS, $\forall i \in \mathcal U$. Referring to \eqref{Eq-RIS-Model2}, we can approximate the reflected signal from the active RIS as follows 
\begin{equation}\label{Eq-RISSig}
    \bm r \approx \underbrace{(\bm I + \bm \Phi \bm H) \bm \Phi}_{\bm \Psi} \left(\sum\limits_{i=1}^U \bm h_{r,i} b_i s_i + \bm z_A\right),
\end{equation}
where $b_i$, $\forall i \in \mathcal U$, is the transmit equalization coefficient of edge node $i$. Given $\bm r$, we can then express the received signal at the edge server as
\begin{align}\label{Eq-RxSig}
    \bm y & = \sum\limits_{i=1}^U \bm h_{d, i} b_i s_i + \bm G \bm r + \bm z_E \nonumber \\[1ex]
    & = \sum\limits_{i=1}^U (\bm h_{d, i} + \bm G \bm \Psi \bm h_{r,i}) b_i s_i + \bm G \bm \Psi \bm z_A + \bm z_E,
\end{align}
where $\bm h_{d,i} \in \mathcal C^{M \times 1}$ denotes the channel from edge node $i$ to the edge server, $\forall i \in \mathcal U$, $\bm G \in \mathcal C^{M \times N}$ denotes the channel from the active RIS to the edge server, and $\bm z_E \sim \left(\bm 0, \sigma_E^2 \bm I \right)$ denotes the thermal noise at the edge server. Note that in \eqref{Eq-RxSig}, we have implicitly assumed that the edge server is equipped with $M \ge 1$ antennas.

By denoting the receive beamforming vector at the edge server as $\bm m \in \mathcal C^{M \times 1}$, we have
\begin{align}\label{Eq-RxSig2}
    \hat s & = \bm m^H \bm y \nonumber \\[2ex]
    & = \bm m^H \sum\limits_{i=1}^U \bm h_{e,i} b_i s_i + \bm m^H \left(\bm G \bm \Psi \bm z_A + \bm z_E\right),
\end{align}
where $\bm h_{e,i} = \bm h_{d, i} + \bm G \bm \Psi \bm h_{r,i}$ is the equivalent channel between edge node $i$ and the edge server, $\forall i \in \mathcal U$.

Note that $\hat s$ in \eqref{Eq-RxSig2} serves as an estimate for $s = \sum\nolimits_{i=1}^U K_i \delta_i s_i$, i.e., the typical entry associated with $\bm s$. However, due to the presence of wireless fading and noise, $\hat s$ does not necessarily equal to $s$. We employ MSE to characterize the distortion between $\hat s$ and $s$, defined as 
\begin{equation}\label{Eq-MSE}
    \mathbb{MSE}(\hat s, s) = \mathbb E\left(|\hat s - s|^2\right).
\end{equation}
Upon obtaining $\hat s$, we use it to recover $g$, i.e., the typical entry associated with $\bm g$, via
\begin{equation}
    \hat g = \frac{1}{K} \left(\hat s + \sum\limits_{i=1}^U \bar g_i\right).
\end{equation}
The MSE between $\hat g$ and $g$ is then given by
\begin{equation}\label{Eq-MSE2}
    \mathbb{MSE}(\hat g, g) = \mathbb E\left(|\hat g - g|^2\right) = \frac{\mathbb E\left(|\hat s - s|^2\right)}{K^2}.
\end{equation}

\section{Convergence Analysis and Problem Formulation}\label{Section-Convergence}
This section analyzes the convergence property of the considered wireless FL system, which motivates the proposed transceiver and RIS configuration design, as detailed below.

\subsection{Convergence Analysis}
To proceed, we first make the following two standard assumptions.
\begin{assumption}
The loss function $L(\cdot)$ is uniformly Lipschitz continuous with parameter $\rho > 0$, such that for any $\bm w, \bm w' \in \mathcal R^{d \times 1}$, we have
\begin{equation}\label{Eq-SmoothAssump}
    L(\bm w') \le L(\bm w) + \left(\bm w' - \bm w\right)^{T} \nabla L(\bm w) + \frac{\rho}{2} \left\|\bm w' - \bm w \right\|^2.
\end{equation}
\end{assumption}

\begin{assumption}
The loss function $L(\cdot)$ is strongly convex with respect to the parameter $\mu > 0$, such that for any $\bm w, \bm w' \in {\cal R}^{d \times 1}$, we have
\begin{equation}\label{Eq-ConvexAssump}
    L(\bm w') \ge L(\bm w) + \left(\bm w' - \bm w\right)^{T} \nabla L(\bm w) + \frac{\mu}{2} \left\|\bm w' - \bm w \right\|^2.
\end{equation}
\end{assumption}

Based on the above two assumptions, we have the following theorem.
\begin{theorem}\label{Theorem-FL-Convergence}
    \emph{Suppose Assumptions 1 and 2 are valid and the learning rate $\eta$ is set to be ${\rho}^{-1}$ for each training round. Then, after $T \ge 1$ training rounds, the expected difference between the training loss $L(\bm w^{[T+1]})$ and the optimal loss $L(\bm w^{\star})$ can be upper bounded by}
    \begin{align}\label{Eq-FL-Convergence}
        {\mathbb E}[L(\bm w^{[T+1]}) - L(\bm w^{\star})] \le {\mathbb E}[L(\bm w^{[1]}) - L(\bm w^{\star})]~\lambda^T \nonumber \\[2ex]
        + \sum\limits_{t=1}^{T} \frac{\lambda^{T-t}}{2 \rho} \mathbb{E}(\|\bm {\hat g}^{[t]} - \bm g^{[t]}\|^2),
    \end{align}
    \emph{where $\bm w^{\star}$ denotes the optimal model parameter vector and $\lambda \triangleq 1 - \mu / \rho$.}
\end{theorem}
\begin{IEEEproof}
    \emph{Refer to Appendix \ref{Theorem-FL-Convergence-Proof}.}
\end{IEEEproof}

Moreover, since $\mu < \rho$, which implies $0 < \lambda < 1$, when $T \to \infty$, $\lambda^T \to 0$, and we can therefore simplify \eqref{Eq-FL-Convergence} as follows
\begin{equation}\label{Eq-FL-Convergence2}
    {\mathbb E}[L(\bm w^{[T+1]}) - L(\bm w^{\star})] \le \sum\limits_{t=1}^{T} \frac{\lambda^{T-t}}{2 \rho} \mathbb{E}(\|\bm {\hat g}^{[t]} - \bm g^{[t]}\|^2).
\end{equation}
It can be observed from \eqref{Eq-FL-Convergence2} that FL recursions over wireless channels still converge, though a gap a,between $L(\bm w^{\star})$ and $\lim\nolimits_{T \to \infty}{\mathbb E}[L(\bm w^{[T+1]})]$ exists due to gradient errors.

\subsection{Problem Formulation}
To improve the performance of the considered wireless FL system, as shown in \eqref{Eq-FL-Convergence2}, we need to minimize the gradient errors in each training round. To this end, we construct the following optimization problem:
\begin{subequations}\label{Eq-OP0}
    \begin{align}
        \min\limits_{\bm m, \bm b, \bm \Phi} ~& \mathbb{MSE}(\hat s, s) \label{OP0-Obj} \\[2ex]
        \text{s.t.}~~ & |b_i|^2 \le P_i, ~\forall i \in \mathcal U, \label{OP0-TxPower} \\[2ex]
        & \mathbb{E}(\|\bm r\|^2) \le P_A, \label{OP0-RISPower} 
\end{align}
\end{subequations}
where $\bm b = [b_1, \cdots, b_U]^T$, and \eqref{OP0-TxPower}, \eqref{OP0-RISPower} account for the maximum power constraints for each edge node and the active RIS, respectively, with $P_i$, $\forall i \in \mathcal U$ denoting the maximum power of edge node $i$, and $P_A$ denoting that of the active RIS. Moreover, we employ $\mathbb{MSE}(\hat s, s)$ instead of $\mathbb{MSE}(\hat g, g)$ as the objective function since minimizing $\mathbb{MSE}(\hat g, \tilde g)$ is equivalent to minimizing $\mathbb{MSE}(\hat s, \tilde s)$, as shown in \eqref{Eq-MSE2}.

\section{Alternative Optimization for Transceiver and RIS Configuration Design}\label{Section-TRDesign}
To proceed, we follow the existing literature \cite{ShuaiWang-TWC, ZhibinWang-TWC, HangLiu-TWC} and assume $\{\bm s_i\}$ are independent of each other, such that both \eqref{OP0-Obj} and \eqref{OP0-RISPower} will possess a closed-form expression. Specifically, when $\mathbb{E}(s_i s_j) = 0$, $\forall i \ne j$, we have
\begin{align}\label{Eq-MSE3}
    & \mathbb{MSE}(\hat s, s) \\[2ex]
    & = \sum\limits_{i=1}^U \left|\bm m^H \bm h_{e,i} b_i - K_i \delta_i \right|^2 + \sigma_A^2 \|\bm m^H \bm G \bm \Psi\|^2 + \sigma_E^2 \|\bm m\|^2, \nonumber
\end{align}
\begin{align}\label{Eq-RISPower}
    \mathbb{E}(\|\bm r\|^2) = \sum\limits_{i=1}^U |b_i|^2 \|\bm \Psi \bm h_{r,i}\|^2 + \sigma_A^2 {\rm Tr}(\bm \Psi \bm \Psi^H) \le P_A.
\end{align}

With \eqref{Eq-MSE3} and \eqref{Eq-RISPower}, it is still challenging to solve \eqref{Eq-OP0} due to the coupling among $\bm m$, $\bm b$, and $\bm \Phi$. In the sequel, we resort to the AO technique to address this issue, which only optimizes one variable at a time, as detailed below.

1) Optimization of $\bm m$: The associated optimization problem with respect to $\bm m$ is given by
\begin{align}\label{OP-m}
    \min\limits_{\bm m} f_0(\bm m) \triangleq & \sum\limits_{i=1}^U \left|\bm m^H \bm h_{e,i} b_i - K_i \delta_i \right|^2 \\[2ex]
    & + \sigma_A^2 \|\bm m^H \bm G \bm \Psi\|^2 + \sigma_E^2 \|\bm m\|^2 \nonumber
\end{align}
which is a least squares problem. The optimal $\bm m$ to \eqref{OP-m} can be found by setting ${\partial f_0(\bm m)}/{\partial \bm m^*}$ to zero, i.e.,
\begin{align}\label{OP-m2}
    \frac{\partial f_0(\bm m)}{\partial \bm m^*} = \bm R \bm m - \sum\limits_{i=1}^U \bm h_{e,i} b_i K_i \delta_i = \bm 0,
\end{align}
which yields
\begin{equation}\label{Optimal-m}
    \bm m^{\star} = \bm R^{-1} \sum\limits_{i=1}^U \bm h_{e,i} b_i K_i \delta_i,
\end{equation}
where $\bm R = \sum\nolimits_{i=1}^U |b_i|^2 \bm h_{e,i} \bm h^H_{e,i} + \sigma_A^2 \bm G \bm \Psi \bm \Psi^H \bm G^H + \sigma_E^2 \bm I$.

2) Optimization of $\bm b$: The associated optimization problem with respect to $\bm b$ is formulated as follows
\begin{subequations}\label{OP-b}
    \begin{align}
        \min\limits_{\bm b} &~ \sum\limits_{i=1}^U \left|\bm m^H \bm h_{e,i} b_i - K_i \delta_i \right|^2 \\[2ex]
        \text{s.t.} &~~ |b_i|^2 \le P_i, ~\forall i \in \mathcal U, \\[2ex]
        &~ \sum\limits_{i=1}^U |b_i|^2 \|\bm \Psi \bm h_{r,i}\|^2 + \sigma_A^2 {\rm Tr}(\bm \Psi \bm \Psi^H) \le P_A.
    \end{align}
\end{subequations}
It is observed that \eqref{OP-b} is a quadratically constrained quadratic program (QCQP), and off-the-shelf solvers such as CVX can be used to solve this problem optimally.

3) Optimization of $\bm \Phi$: The associated optimization problem with respect to $\bm \Phi$ is formulated as follows
\begin{subequations}\label{OP-Phi}
\begin{align}
    \min\limits_{\bm \Phi} &~ f_1(\bm \Phi) \\[2ex]
    \text{s.t.} &~ g_1(\bm \Phi) \le P_A,
\end{align}
\end{subequations}
where $f_1(\bm \Phi)$ and $g_1(\bm \Phi)$ are respectively given by
\begin{align*}
    f_1(\bm \Phi) = & \sum\limits_{i=1}^U \left|\bm m^H (\bm h_{d,i} + \bm G (\bm I + \bm \Phi \bm H) \bm \Phi \bm h_{r,i}) b_i - K_i \delta_i\right|^2 \\[2ex]
    & + \sigma_A^2 \|\bm m^H \bm G (\bm I + \bm \Phi \bm H) \bm \Phi\|^2, \\[2ex]
    g_1(\bm \Phi) = & \sum\limits_{i=1}^U |b_i|^2 \|(\bm I + \bm \Phi \bm H) \bm \Phi \bm h_{r,i}\|^2 \\[2ex]
    & + \sigma_A^2 {\rm Tr}((\bm I + \bm \Phi \bm H) \bm \Phi \bm \Phi^H (\bm I + \bm \Phi \bm H)^H).
\end{align*}
To handle \eqref{OP-Phi}, we introduce an auxiliary variable $\bm {\tilde \Phi}$ and reformulate \eqref{OP-Phi} as follows
\begin{subequations}\label{OP-Phi2}
\begin{align}
    \min\limits_{\bm \Phi, \bm {\tilde \Phi}} &~ f_2(\bm \Phi, \bm {\tilde \Phi}) + \tau \|\bm \Phi - \bm {\tilde \Phi}\|_{\rm F}^2 \\[2ex]
    \text{s.t.} &~ g_2(\bm \Phi, \bm {\tilde \Phi}) \le P_A,
\end{align}
\end{subequations}
where $f_2(\bm \Phi, \bm {\tilde \Phi})$ and $g_2(\bm \Phi, \bm {\tilde \Phi})$ are respectively given by
\begin{align*}
    f_2(\bm \Phi, \bm {\tilde \Phi}) = & \sum\limits_{i=1}^U \left|\bm m^H (\bm h_{d,i} + \bm G (\bm I + \bm {\tilde \Phi} \bm H) \bm \Phi \bm h_{r,i}) b_i - K_i \delta_i\right|^2 \\[2ex]
    & + \sigma_A^2 \|\bm m^H \bm G (\bm I + \bm {\tilde \Phi} \bm H) \bm \Phi\|^2, \\[2ex]
    g_2(\bm \Phi, \bm {\tilde \Phi}) = & \sum\limits_{i=1}^U |b_i|^2 \|(\bm I + \bm {\tilde \Phi} \bm H) \bm \Phi \bm h_{r,i}\|^2 \\[2ex]
    & + \sigma_A^2 {\rm Tr}((\bm I + \bm {\tilde \Phi} \bm H) \bm \Phi \bm \Phi^H (\bm I + \bm {\tilde \Phi} \bm H)^H).
\end{align*}
Note that $\tau$ in \eqref{OP-Phi2} is a penalty parameter, and it can be proven that \eqref{OP-Phi2} is equivalent to \eqref{OP-Phi} when $\tau \to \infty$. In the sequel, we recall the AO technique and optimize $\bm \phi \triangleq {\rm diag}(\bm \Phi)$ and $\bm {\tilde \phi} \triangleq {\rm diag}(\bm {\tilde \Phi})$ alternatively until $\bm \phi = \bm {\tilde \phi}$ is achieved.

3$a$) Optimization of $\bm \phi$: Through some mathematical manipulations to \eqref{OP-Phi2}, we formulate an optimization problem with respect to $\bm \phi$ as follows
\begin{subequations}\label{OP-Phi3}
\begin{align}
    \min\limits_{\bm \phi} &~ \bm \phi^H \bm A_1 \bm \phi - \bm \phi^H \bm v_1 - \bm v_1^H \bm \phi \\[2ex]
    \text{s.t.} &~ \bm \phi^H \bm B_1 \bm \phi \le P_A,
\end{align}
\end{subequations}
where $\bm A_1$, $\bm B_1$, and $\bm v_1$ are respectively given by
\begin{align*}
    & \bm A_1 = \sum\limits_{i=1}^U \bm a_i \bm a_i^H + \tau \bm I \\[2ex]
    &~~~~~~~ + \sigma_A^2 {\rm diag}(\bm m^H \bm G \bm \Omega) {\rm diag}(\bm \Omega^H \bm G^H \bm m), \\[2ex]
    & \bm B_1 = \sum\limits_{i=1}^U |b_i|^2 {\rm diag}(\bm h_{r,i}^*) \bm \Omega^H \bm \Omega {\rm diag}(\bm h_{r,i}) \\[2ex]
    &~~~~~~~ + \sigma_A^2 ((\bm \Omega^H \bm \Omega) \odot \bm I), \\[2ex]
    & \bm \Omega = \bm I + \bm {\tilde \Phi} \bm H, \\[2ex]
    & \bm v_1 = \sum\limits_{i=1}^U \left(K_i \delta_i - \bm m^H \bm h_{d,i} b_i\right) \bm a_i + \tau \bm {\tilde \phi}, \\[2ex]
    & \bm a_i = {\rm diag}(\bm h_{r,i}^* b_i^*) \bm \Omega^H \bm G^H \bm m,~\forall i \in \mathcal U.
\end{align*}
It is observed that \eqref{OP-Phi3} is a standard QCQP, which can be solved optimally using the Karush-Kuhn-Tucker (KKT) conditions, as detailed below.

First of all, the Lagrangian associated with \eqref{OP-Phi3} is defined as follows
\begin{equation}\label{OP-Phi3-Lagrangian}
    F_1 = \bm \phi^H \bm A_1 \bm \phi - \bm \phi^H \bm v_1 - \bm v_1^H \bm \phi + \lambda_1 (\bm \phi^H \bm B_1 \bm \phi - P_A),
\end{equation}
where $\lambda_1 \ge 0$ is the Lagrange multiplier. The KKT conditions of \eqref{OP-Phi3-Lagrangian} are then given by
\begin{subequations}
    \begin{align}
        \frac{\partial F_1}{\partial \bm \phi^*} = \left(\bm A_1 + \lambda_1 \bm B_0 \right) \bm \phi - \bm v_1 = \bm 0, \label{OP-Phi3-KKT-phi} \\[2ex]
        \lambda_1 (\bm \phi^H \bm B_1 \bm \phi - P_A) = 0, \label{OP-Phi3-KKT-Slackness} \\[2ex]
        \bm \phi^H \bm B_1 \bm \phi = P_A. \label{OP-Phi3-KKT-PowerCons}
     \end{align}
\end{subequations}
From \eqref{OP-Phi3-KKT-phi}, we can compute the optimal $\bm \phi$ as
\begin{equation}\label{Optimal-phi}
    \bm \phi^{\star} = \left(\bm A_1 + \lambda_1 \bm B_1 \right)^{-1} \bm v_1,
\end{equation}
where the nonnegative Lagrange multiplier $\lambda_1$ should be chosen to satisfy \eqref{OP-Phi3-KKT-Slackness} and \eqref{OP-Phi3-KKT-PowerCons}. With \eqref{Optimal-phi}, we can verify that $(\bm \phi^{\star})^H \bm B_1 \bm \phi^{\star}$ is a decreasing function of $\lambda_1$. Moreover, it can be shown that
\begin{equation}\label{lambda-bounds}
    0 \le \lambda_1 < \sqrt{\frac{\bm v_1^H \bm B_1^{-1} \bm v_1}{P_A}}.
\end{equation}
Therefore, we can search for $\lambda_1$ using the bisection search method within the bounds on $\lambda_1$ in \eqref{lambda-bounds}.

3$b$) Optimization of $\bm {\tilde \phi}$: The assocaited optimization problem with respect to $\bm {\tilde \phi}$ is formulated as follows
\begin{subequations}\label{OP-Phi4}
\begin{align}
    \min\limits_{\bm {\tilde \phi}} &~ \bm {\tilde \phi}^H \bm A_2 \bm {\tilde \phi} - \bm {\tilde \phi}^H \bm v_2 - \bm v_2^H \bm {\tilde \phi} \\[2ex]
    \text{s.t.} &~ \bm {\tilde \phi}^H \bm B_2 \bm {\tilde \phi} + \bm q^H \bm {\tilde \phi} + \bm {\tilde \phi}^H \bm q + {\rm Tr}(\bm D) \le P_A,
\end{align}
\end{subequations}
where $\bm A_2$, $\bm B_2$, $\bm D$, $\bm v_2$, and $\bm q$ are respectively given by
\begin{align*}
    & \bm A_2 = \sum\limits_{i=1}^U \bm {\tilde a}_i \bm {\tilde a}_i^H + \tau \mathbf I \nonumber \\[2ex]
    &~~~~~~~ + \sigma_A^2 {\rm diag}(\bm G^H \bm m) \bm H^* \bm \Phi^H \bm \Phi \bm H^T {\rm diag}(\bm m^H \bm G), \\[2ex]
    & \bm B_2 =  (\bm H \bm D \bm H^H) \odot \bm I, \\[2ex]
    & \bm D =  \bm \Phi \left(\sum\limits_{i=1}^U |b_i|^2 \bm h_{r,i} \bm h^H_{r,i} + \sigma_A^2 \bm I\right) \bm \Phi^H, \\[2ex]
    & \bm v_2 = \sum\limits_{i=1}^U \left[K_i \delta_i - \bm m^H (\bm h_{d,i} + \bm G \bm \Phi \bm h_{r,i}) b_i\right] \bm {\tilde a}_i + \tau {\bm \phi} \\[2ex]
    &~~~~~~ - \sigma_A^2 {\rm diag}(\bm m^H \bm G \bm \Phi \bm \Phi^H \bm H^H) \bm G^H \bm m, \\[2ex]
    & \bm q = {\rm diag}(\bm D \bm H^H), \\[2ex]
    & \bm {\tilde a}_i = {\rm diag}(\bm H^* \bm \Phi^* \bm h^*_{r,i} b_i^*) \bm G^H \bm m, ~\forall i \in \mathcal U.
\end{align*}
It is observed that \eqref{OP-Phi4} is a QCQP, and we can also employ the KKT conditions to solve it, as detailed below.

First of all, the Lagrangian associated with \eqref{OP-Phi4} is expressed as follows
\begin{align}\label{OP-Phi4-Lagrangian}
    F_2 = &~ \bm {\tilde \phi}^H \bm A_2 \bm {\tilde \phi} - \bm {\tilde \phi}^H \bm v_2 - \bm v_2^H \bm {\tilde \phi} \nonumber \\[2ex]
    &~ + \lambda_2 (\bm {\tilde \phi}^H \bm B_2 \bm {\tilde \phi} + \bm q^H \bm {\tilde \phi} + \bm {\tilde \phi}^H \bm q + {\rm Tr}(\bm D) - P_A),
\end{align}
where $\lambda_2 \ge 0$ is the Lagrange multiplier. The KKT conditions of \eqref{OP-Phi4-Lagrangian} are given by
\begin{subequations}
    \begin{align}
        \frac{\partial F_2}{\partial \bm {\tilde \phi}^*} = \left(\bm A_2 + \lambda_2 \bm B_2 \right) \bm {\tilde \phi} - \bm v_2 + \lambda_2 \bm q = \bm 0, \label{OP-Phi4-KKT-phi} \\[2ex]
        \lambda_2 (\bm {\tilde \phi}^H \bm B_2 \bm {\tilde \phi} + \bm q^H \bm {\tilde \phi} + \bm {\tilde \phi}^H \bm q + {\rm Tr}(\bm D) - P_A) = 0, \label{OP-Phi4-KKT-Slackness} \\[2ex]
        \bm {\tilde \phi}^H \bm B_2 \bm {\tilde \phi} + \bm q^H \bm {\tilde \phi} + \bm {\tilde \phi}^H \bm q + {\rm Tr}(\bm D) = P_A. \label{OP-Phi4-KKT-PowerCons}
     \end{align}
\end{subequations}
From \eqref{OP-Phi4-KKT-phi}, we can derive the optimal $\bm {\tilde \phi}$ as
\begin{equation}\label{Optimal-phi2}
    \bm {\tilde \phi}^{\star} = \left(\bm A_2 + \lambda_2 \bm B_2 \right)^{-1} (\bm v_2 - \lambda_2 \bm q),
\end{equation}
where the nonnegative Lagrange multiplier $\lambda_2$ can be determined through the one-dimensional grid search to satisfy \eqref{OP-Phi4-KKT-Slackness} and \eqref{OP-Phi4-KKT-PowerCons}.

Thus far, we have introduced the proposed transceiver and RIS configuration design approach and the whole procedures are summarized in \textbf{Algorithm \ref{Alg-TRDesign}} for clarity.

\begin{algorithm}[!htbp]
    \caption{Pseudo-Code for the Proposed Transceiver and RIS Configuration Design Approach}
    \begin{algorithmic}[1]
        \State Initialize $\bm b$, and $\bm \Phi$ to ensure that the power constraints in \eqref{OP0-TxPower} and \eqref{OP0-RISPower} are satisfied.
        \While{not converge}
        \State Update $\bm m$ through \eqref{Optimal-m}.
        \State Update $\bm b$ through solving \eqref{OP-b}.
        \While{$\bm \phi \ne \bm {\tilde \phi}$}
        \State Compute $\bm \phi$ through \eqref{Optimal-phi}.
        \State Compute $\bm {\tilde \phi}$ through \eqref{Optimal-phi2}.
        \State Update $\tau$ by $\tau \leftarrow 1.1 \times \tau$.
        \EndWhile
        \State Update $\bm \Phi$ by $\bm \Phi = {\rm diag}(\bm \phi)$.
        \EndWhile
    \end{algorithmic}\label{Alg-TRDesign}
\end{algorithm}


\section{Numerical Results}\label{Section-NR}
In this section, we present numerical results to demonstrate the effectiveness of deploying active RIS in enhancing the performance of the AirComp-enabled FL system. We adopt a three-dimensional coordinate configuration, where the locations of the edge server and the active RIS are set to $\left(-50, 0, 10\right)$ meters and $\left(0, 0, 10\right)$ meters, respectively, and the $U = 20$ edge nodes are uniformly distributed in the region of $\left([0,20], [-10, 10], 0\right)$ meters. Each link in $\{\bm h_{r,i}\}$, $\{\bm h_{d,i}\}$, and $\bm G$ is subjected to both path loss and small-scale fading. The path loss model is given by $\textsf{PL}(\xi) = C_0 \left(\xi / \xi_0\right)^{-\kappa}$, where $C_0 = 30$ dB denotes the path loss at the reference distance of $\xi_0 = 1$ meter, $\xi$ represents the link distance, and $\kappa$ is the path loss component. Throughout the simulations, the path loss components for $\{\bm h_{r,i}\}$, $\{\bm h_{d,i}\}$, and $\bm G$ are set to $2.8$, $3.6$, and $2.2$, respectively. For small-scale fading, we employ the standard Rician channel model, assigning Rician factors of $0$, $0$, and $3$ dB to $\{\bm h_{r,i}\}$, $\{\bm h_{d,i}\}$, and $\bm G$, respectively. Furthermore, we set $M = 10$, $N = 200$, $P_i = 0$ dB, $\forall i \in \mathcal U$, $P_A = 0$ dB, $\sigma_A^2 = -80$ dB, $\sigma_E^2 = -80$ dB, and $\nu = -30$ dB.

To evaluate learning performance, we use the MNIST dataset to simulate the handwritten digit recognition task \cite{MNIST}. Specifically, we train a fully connected neural network with $784$ inputs and $10$ outputs, using cross-entropy as the loss function, which yields a total of $d = 7840$ model parameters. The set of $K = 60,000$ training data samples is equally divided into $40$ shards of size $1500$ in a non-IID manner, and we assign each edge node two shards without replacement as its local dataset, i.e., $K_i = 3000$, $\forall i \in \mathcal U$. The test dataset consists of $10,000$ samples, and we evaluate learning performance using test accuracy, defined as $\frac{\text{number of correctly recognized handwritten digits}}{10000}$. Moreover, the number of training rounds $T$ is set to be $50$, and the learning rate $\eta^{[t]} = 0.05$, $\forall t = 1, \cdots, T$.

\begin{figure}
\vskip -8pt
\centering
\subfigure[]
{\includegraphics[width = 7.8cm]{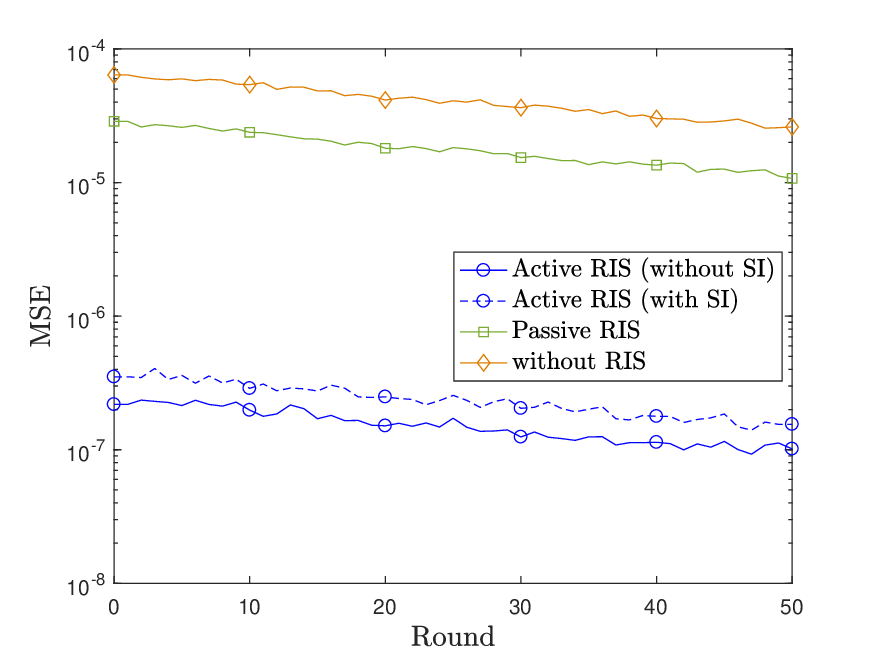} \label{MSE-Round}}
\subfigure[]
{\includegraphics[width = 7.8cm]{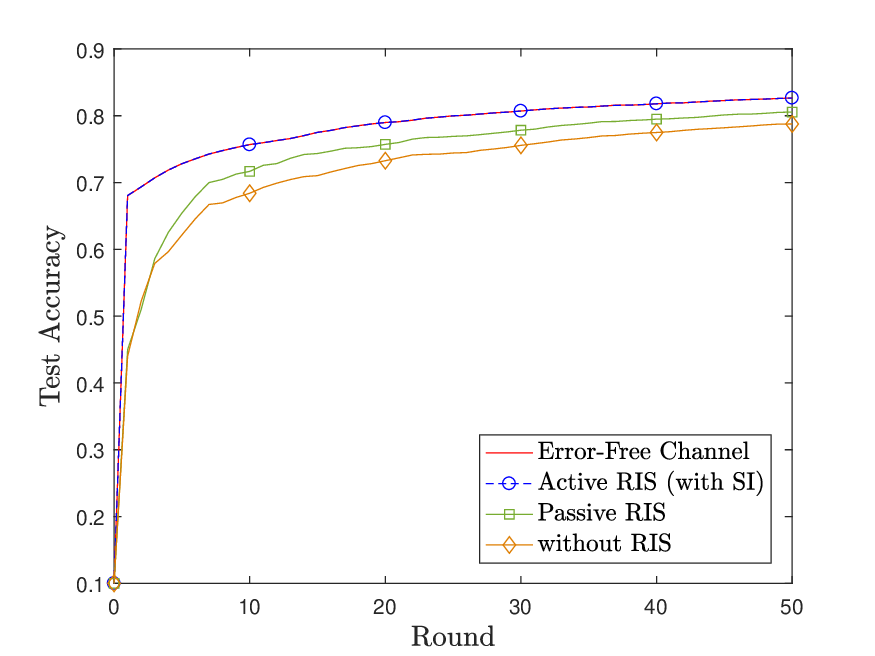} \label{ACC-Round}}
\caption{Gradient aggregation error $(a)$ and test accuracy achieved by current global model parameter vector $(b)$ versus training round.}
\end{figure}

In Fig. \ref{MSE-Round}, we plot $\mathbb{E}\left(\|\bm {\hat g}^{[t]} - \bm g^{[t]}\|^2\right) / d$, $\forall t = 1, \cdots, T$. From this figure, it is immediately apparent that the active RIS can significantly reduce the errors in gradient aggregation compared to FL without RIS, whereas the passive RIS achieves only limited improvement. Consequently, we can deduce that the active RIS is more efficient than its passive counterpart in enhancing the performance of the AirComp-enabled FL system, as corroborated by Fig. \ref{ACC-Round}.

\section{Conclusions}\label{Section-CN}
In this paper, we have proposed an active RIS assisted AirComp technique to support gradient aggregation in wireless FL systems. Our analysis of the FL convergence property has revealed that minimizing gradient aggregation errors in each training round is pivotal to reducing the convergence gap. As such, we have developed an AO approach for the joint optimization of the transceiver design and RIS configuration in each training round. Experiment results have demonstrated that the active RIS significantly reduced gradient aggregation errors compared to its passive counterpart, thereby leading to superior learning performance.

\begin{appendices}
\section{}\label{Theorem-FL-Convergence-Proof}
Referring to \cite{MPFriedlander}, when the loss function $L(\bm w)$ is uniformly Lipschitz continuous with parameter $\rho$ and the learning rate $\eta^{[t]} = {1}/{\rho}$, the following inequality holds:
\begin{equation}\label{Eq-OneIter}
    L(\bm w^{[t+1]}) \le L(\bm w^{[t]}) - \frac{1}{2\rho}\|\nabla L(\bm w^{[t]})\|^2 + \frac{1}{2\rho}\|\bm e^{[t]}\|^2,
\end{equation}
where $\bm e^{[t]} = \bm {\hat g}^{[t]} - \bm g^{[t]}$ is the gradient error. Moreover, from \eqref{Eq-ConvexAssump}, we can derive that
\begin{equation}\label{Eq-Grad-LB}
    \|\nabla L(\bm w^{[t]})\|^2 \ge 2 \mu [L(\bm w^{[t]}) - L(\bm w^{\star})].
\end{equation}
By substituting \eqref{Eq-Grad-LB} into \eqref{Eq-OneIter}, we have
\begin{align}\label{Eq-OneIter2}
    & L(\bm w^{[t+1]}) \nonumber \\[2ex] 
    & ~~ \le L(\bm w^{[t]}) - ({\mu}/{\rho})~ [L(\bm w^{[t]}) - L(\bm w^{\star})] + \frac{1}{2\rho} \|\bm e^{[t]}\|^2.
\end{align}
Next, by first subtracting $L(\bm w^{\star})$ and then taking expectation on both sides of \eqref{Eq-OneIter2}, we obtain that
\begin{align}\label{Eq-OneIter3}
    {\mathbb E}[L(\bm w^{[t+1]}) - L(\bm w^{\star})] ~~~~~~~~~~~~~~~~~~~~~~~~~~~~~~~~~~~~~ \nonumber \\[2ex]
    \le (1 - \mu/\rho)~{\mathbb E}[L(\bm w^{[t]}) - L(\bm w^{\star})] + \frac{1}{2\rho} {\mathbb E}[\|\bm e^{[t]}\|^2].
\end{align}
Applying \eqref{Eq-OneIter3} recursively for $t = T, \cdots, 1$, we prove \eqref{Eq-FL-Convergence} and complete the proof.
\end{appendices}

\end{document}